\documentstyle[11pt]{article}
\def\dif{{\rm d}}
\def\deriv{\@ifnextchar[{\@deriv}{\@deriv[]}}
   \def\@deriv[#1]#2#3{\mathchoice%
{{\dif^{#1}#2\over\dif{#3}^{#1}}}{{\dif^{#1}#2/\dif{#3}^{#1}}}%
+{{\dif^{#1}#2\over\dif{#3}^{#1}}}{{\dif^{#1}#2/\dif{#3}^{#1}}}}

\def\secteqno{\@addtoreset{equation}{section}%
\def\theequation{\thesection.\arabic{equation}}}
\newcounter{subequation}
\def\thesubequation{\alph{subequation}}
\def\sneqnarray{\stepcounter{equation}\let\@currentlabel=\theequation
\setcounter{subequation}{1}
\def\@eqnnum{{\rm (\theequation.\thesubequation)}}
\global\@eqcnt\z@\tabskip\@centering\let\\=\@eqncr\let\@@eqncr=\@@sne
qncr
$$\halign to \displaywidth\bgroup\@eqnsel\hskip\@centering
 $\displaystyle\tabskip\z@{##}$&\global\@eqcnt\@ne
 \hskip 2\arraycolsep \hfil${##}$\hfil
 &\global\@eqcnt\tw@ \hskip 2\arraycolsep $\displaystyle\tabskip\z@{##}$\hfil
  \tabskip\@centering&\llap{##}\tabskip\z@\cr}
\def\endsneqnarray{\@@sneqncr\egroup $$\global\@ignoretrue}
\def\@@sneqncr{\let\@tempa\relax
   \ifcase\@eqcnt \def\@tempa{& & &}\or \def\@tempa{& &}
   \else \def\@tempa{&}\fi
     \@tempa \if@eqnsw\@eqnnum\stepcounter{subequation}\fi
     \global\@eqnswtrue\global\@eqcnt\z@\cr}
\def\nobiblabels{\def\@lbibitem[##1]##2{\@bibitem{##2}}}
\makeatother
\def\ben{\begin{enumerate}}
\def\een{\end{enumerate}}
\def\beq{\begin{equation}}
\def\eeq{\end{equation}}
\def\bea{\begin{eqnarray}}
\def\eea{\end{eqnarray}}
\def\beann{\begin{eqnarray*}}
\def\eeann{\end{eqnarray*}}
\def\beasn{\begin{sneqnarray}}
\def\eeasn{\end{sneqnarray}}

\newtheorem{teor}{Theorem}

\def\sota#1{\mathrel{\mathop{=}\limits_{#1}}}
\begin{document}

\title{Lagrangian Noether symmetries as  canonical
transformations}

\author{J.\ Antonio \ Garc\'{\i}a
\thanks{Electronic address: garcia@nuclecu.unam.mx}\\
Instituto de Ciencias Nucleares\\
   Universidad Nacional Aut\'onoma de M\'exico\\
   Apartado Postal 70-543, 04510 \\
   M\'exico, D.F.\\
and
\\
J.\ M.\ Pons
\thanks{Electronic address: pons@ecm.ub.es}\\
Departament d'Estructura i Constituents de la Mat\`eria,\\ 
Universitat de Barcelona,\\
and
\\
 Institut de F\'\i sica d'Altes Energies,\\
Diagonal 647, E-08028 Barcelona, Catalonia, Spain}
%\\
% and\\
%   Center for Relativity, Physics Department, \\
%   The University of Texas, Austin, Texas 78712-1081, USA}

\maketitle

%234567890 234567890 234567890 234567890 234567890 234567890 234567890

%%%%%%%%%%%%%%%%%%%%%%%%%%%%%%%%%%%%%%%%%%%%%%%%%%%%%%%%%%%%%%%%
%%%%%%%%%%%%%%%%%%%%%%%%%%%%%%%%%%%%%%%%%%%%%%%%%%%%%%%%%%%%%%%%

\begin{abstract}
We prove that, given a time-independent Lagrangian defined in the first 
tangent bundle of configuration space, every infinitesimal Noether 
symmetry that is defined in the $n$-tangent bundle and is  not vanishing 
on-shell, can be written as a canonical symmetry
in an enlarged phase space, up to constraints that vanish on-shell. 
The proof is performed by the implementation 
of a change of variables from the the $n$-tangent bundle 
of the Lagrangian theory to an extension of the Hamiltonian 
formalism which is particularly suited for the case when the Lagrangian is 
singular. This result proves the assertion that any Noether symmetry can 
be canonically realized in an enlarged phase space.
Then we work out the regular case as a particular application of this ideas
and rederive the
Noether identities in this framework. Finally we present an example 
to illustrate our results.
\end{abstract}

%\pacs{11.10.Ef, 11.30.-j, 04.20.Fy  \hfill hep-th/?????}

\catcode`\@=11
\newdimen\ex@
\ex@.2326ex
\def\dddot#1{{\mathop{#1}\limits^{\vbox to-1.4\ex@{\kern-\tw@\ex@
\hbox{\rm...}\vss}}}}
\catcode`\@=\active

%%%%%%%%%%%%%%%%%%%%%%%%%%%%%%%%%%%%%%%%%%%%%%%%%%%%%%%%%%%%%%%%%%%%%%

\section{Introduction.}

\subsection{Noether transformations}

In gauge theories, the conserved quantities associated with 
infinitesimal Noether transformations exhibit some features that are absent
in a regular (non-gauge) theory. In \cite{noeth98} we have studied 
exhaustively the characterization of such conserved quantities in some
specific contexts. We always consider as our starting point a time-independent
first order Lagrangian $L(q,\, \dot q)$ defined in configuration-velocity 
space $TQ$, 
that is, the tangent bundle of some configuration manifold $Q$.
Gauge theories rely on Lagrangians whose 
Hessian matrix with respect to the velocities ($q$ stand 
for local coordinates in $Q$ \footnote{In most of the paper we will 
ignore the coordinate labels for $q$, that is, we will often write $q$ 
instead of $q^i$.})
\beq
W_{ij}\equiv {\partial^2L\over\partial\dot q^i\partial\dot q^j},
\label{hess}
\eeq 
is not invertible (singular Lagrangians); this fact implies the 
existence of constraints in phase space. The canonical
treatment of these theories was first solved by Dirac \cite{dirac}. 

For the $n$-th 
tangent bundle of $Q$ (we leave $n$ undetermined but finite in order 
to keep locality),
the most general infinitesimal Noether transformations that can be constructed
for either a singular or regular Lagrangian $L(q,\dot q)$, is a set of 
functions of the form 
\beq
\delta^L q^i (q,\, \dot q, \ddot q, ... ;\,t),
\label{gen-delq}
\eeq
where $q,\, \dot q, \ddot q, ...$ are the local coordinates in the $n$-th 
tangent bundle. 
These functions are
such that they induce a transformation $\delta L$ satisfying
\beq
\delta L = \frac{d^L\!F}{d\,t},
\label{dell}
\eeq
for some infinitesimal function $F(q,\, \dot q, \ddot q, ... ;\,t)$. 
Property (\ref{dell}) completely characterizes the Noether symmetries. 
The total 
time derivative operator in the $n$ tangent bundle is defined as 
\beq
\frac{d^L}{d\,t} = \frac{\partial}{\partial\,t} + 
{\dot q^i}\frac{\partial}{\partial\,q^i} +
{\ddot q^i}\frac{\partial}{\partial\, {\dot q^i}} + ... \ .
\label{timeder}
\eeq

Equation (\ref{dell}) can be equivalently written as 
\begin{equation} 
[L]_i\delta^L q^i + \frac{d^L G^L}{d \, t} = 0,
\label{noet}
\end{equation}
where $[L]_i$ stands for the Euler-Lagrange equations
$$
[L]_i := \alpha_i - W_{is}\ddot q^s,
$$
with
$$
\alpha_i :=
    - {\partial^2L\over\partial\dot q^i\partial q^s}\dot q^s
    + {\partial L\over\partial q^i} .
$$
The conserved quantity is then 
$
G^L = ({\partial L}/{\partial \dot q^i}) \delta q^i - F.
$
In a gauge theory the infinitesimal Noether 
transformations (\ref{gen-delq}) may
contain arbitrary infinitesimal functions; these are the 
Noether gauge transformations. In such case the associated 
conserved quantity is zero on-shell, that is, it is a combination of 
--first class-- constraints.

%%%%%%%%%%%%%%%%%%%%%%%%%%%%%%%%%%%%%%%%%%%%%%%%%%%%%%%%%%%%%%%%%%%%%%%%
%%%%%%%%%%%%%%%%%%%%%%%%%%%%%%%%%%%%%%%%%%%%%%%%%%%%%%%%%%%%%%%%%%%%%%%%

%%%%%%%%%%%%%%%%%%%%%%%%%%%%%%%%%%%%%%%%%%%%%%%%%%%%%%%%%%%%%%%%%%%%%%%%
%%%%%%%%%%%%%%%%%%%%%%%%%%%%%%%%%%%%%%%%%%%%%%%%%%%%%%%%%%%%%%%%%%%%%%%%

\subsection{Noether transformations in the enlarged formalism.}

The enlarged formalism is a Lagrangian treatment of the Hamilton variational
principle which is particularly convenient for singular Lagrangians.
This formalism takes as a starting point the canonical Lagrangian $L_c$. 
Once a canonical hamiltonian $H_c$ and a complete set 
of independent primary constraints $\phi_\mu$ are determined out of 
the original Lagrangian $L$, $L_c$
is defined as follows, 
\beq
L_c(q,p,\lambda;\dot q, \dot p, {\dot \lambda})
:= p {\dot q} - H_c(q,p) - \lambda^\mu \phi_\mu(q,p).
\label{canlagr}
\eeq
The new configuration space for $L_c$ is the old phase space enlarged with
the Lagrange multipliers $\lambda^\mu$ as new independent variables.
The dynamics given by $L_c$ is nothing but the constrained Dirac's
Hamiltonian dynamics for a system with canonical Hamiltonian
$H_c$ and a number of primary constraints $\phi_\mu$.

We will consider Noether transformations for $L_c$ that are canonically 
generated for what regards the variables $q$'s and $p$'s, but that can depend
also on the Lagrange multipliers $\lambda^\mu$ and their time 
derivatives at any finite order.
In \cite{noeth98} we have established the condition for a function
$G^c(q,p,\lambda, \dot \lambda, \ddot \lambda,...;t)$ to be a Noether
generator for $L_c$ according to the rules
\beq
\delta_c q = \{q, \, G^c \} \quad
\delta_c p = \{p, \, G^c \}.
\label{delc}
\eeq

The condition is 
\beq
\frac{D \, G^c}{D \, t} + \{ G^c,\, H_D \} = pc,
\label{enl-cond}
\eeq
where $pc$ stands for an arbitrary linear combination of the 
primary constraints $\phi_\mu$; 
$H_D$ is the Dirac Hamiltonian defined by
$$
H_D = H_c + \lambda^\mu \phi_\mu,
$$ 
and we have also introduced the notation
\beq
\frac{D \, }{D \, t} :=
\frac{\partial }{\partial \, t} + {\dot \lambda^\mu} \frac{\partial
}{\partial \, \lambda} + {\ddot \lambda^\mu} \frac{\partial
}{\partial \, {\dot \lambda}} +... \ .
\label{Dt}
\eeq

The right hand side of (\ref{enl-cond}) contains the 
prescription to define 
$\delta_c \lambda$. Indeed, if ``$pc$'' in the right hand side of 
(\ref{enl-cond}) is $C^\mu \phi_\mu$, then $\delta_c \lambda^\mu= C^\mu$. 
We reproduce in the appendix the proof of the condition 
(\ref{enl-cond}) as  given in \cite{noeth98}.

In \cite{htz} an algorithm was introduced
to obtain gauge generators for theories with only first class constraints. One
can easily check that our expression (\ref{enl-cond}) condenses all the
contents of this algorithm. The generality of (\ref{enl-cond}) lies in the 
fact that it extends the formalism of \cite{htz} to a general gauge theory 
-with first and second class constraints- and that it covers the case 
of rigid Noether transformations as well.

One might wonder whether these Noether transformations for the enlarged 
Lagrangian $L_c$ are still Noether transformations for the Lagrangian $L$. 
The answer is in the positive.
It is proved in \cite{noeth98} that we can obtain a 
transformation $\delta^L q$ of the type
(\ref{gen-delq}) and satisfying (\ref{dell}) out of our $\delta_c q$. Indeed,
to go back to the original Lagrangian $L$ and to 
establish its corresponding Noether transformation associated with 
$\delta_c q $ of (\ref{delc}), we must
produce a pullback from phase space to configuration-velocity space 
in order to substitute the momenta $p$ by their Lagrangian
definition ${\hat p} := {\partial L}/{\partial {\dot q}}$, but we must 
also
substitute the variables $\lambda, \dot \lambda, \ddot \lambda,...$ by their
Lagrangian counterparts. This is easily done by noticing that the 
Euler-Lagrange equations of (\ref{canlagr}) for $p$ and $\lambda$ 
can be used to isolate these variables in terms of $q$ and $\dot q$. 
In fact, from the equations	
$$
[L_c]_p = \dot q - 
{\partial H_c\over\partial p} - \lambda^\mu {\partial
\phi_\mu\over\partial p} = 0,
$$
and
$$
[L_c]_{\lambda^\mu} = -\phi_\mu = 0,
$$
we get
$p = {\hat p}(q,\dot q)$ 
and a Lagrangian determination for $\lambda^\mu$ that
defines a new set of functions $v^\mu$ such that 
$\lambda^\mu = v^\mu(q, \dot q)$. These functions $v^\mu(q, \dot q)$, 
with some properties, were already introduced in \cite{batlle86}.
We arrive at the definition for $\delta^L q$ 
\beq
\delta^L q(q,\, \dot q, \ddot q, ... ;\,t) :=
\delta_c q(q,{\hat p},v^\mu, {\dot v^\mu}, {\ddot v^\mu},...;t), 
\label{def-del}
\eeq 
where $\dot v^\mu = {d^L \,v^\mu}/{d \, t}$, etc.. It is proved in 
\cite{noeth98} that this $\delta^L q$ defined in (\ref{def-del}) is indeed 
a Noether transformation for $L$ having
$$
G^L(q,\, \dot q, \ddot q, ... ;\,t) :=
G^c(q,{\hat p},v^\mu, {\dot v^\mu}, {\ddot v^\mu},...;t), 
$$
as its associated conserved quantity.

The scope of the present paper is to show that this type of 
transformations 
(\ref{delc}), generated by functions $G^c$ satisfying (\ref{enl-cond}), is 
general enough to encompass the most general Lagrangian 
Noether transformation of the 
type (\ref{gen-delq}). We will prove that the 
transformations (\ref{def-del})
are, up to the addition of constraints --that vanish on shell--, 
the most general  
transformations satisfying (\ref{dell}), that is, the most general Noether 
transformations. The proof will be based on a change of variables that is 
particularly suited to connect the Hamiltonian and the Lagrangian formalism 
in the singular case. These new variables are introduced in the next section, 
whereas in section 3 we prove the main result of this paper. 
A simple application to the regular case is given in section 4. As a bonus,
in section 5, we rederive in this framework the Noether identities for 
singular Lagrangians and in section 6 we present an example to illustrate our 
result. Finally, we reproduce in the appendix the proof of 
some results quoted in the introduction.

%%%%%%%%%%%%%%%%%%%%%%%%%%%%%%%%%%%%%%%%%%%%%%%%%%%%%%%%%%%%%%%%%%%%%%
%%%%%%%%%%%%%%%%%%%%%%%%%%%%%%%%%%%%%%%%%%%%%%%%%%%%%%%%%%%%%%%%%%%%%%
\section{From the old to the new variables}

As a warm up, let us observe that, in a regular case, that is, when 
(\ref{hess}) is invertible, we can use the equations of motion, 
that are expressible in normal form as 
$\ddot q = W^{-1}\alpha$,
to eliminate the higher order dependences in (\ref{gen-delq}); this elimination
results in a new transformation $\delta^L_0 q$ in $TQ \times R$, given by
$$
\delta^L_0 q (q, \dot q ;t) \equiv \delta^L q 
(q,\, \dot q, \ddot q, ... ;\,t)_{|_{[L]=0}},
$$
where in $[L]=0$ we include its time derivatives at any order that is 
needed.

Now the question: is $\delta^L_0 q$ a Noether transformation for $L$?
The answer is yes and the proof will come out
as a particular case of the general results we will obtain below.

%%%%%%%%%%%%%%%%%%%%%%%%%%%%%%%%%%%%%%%%%%%%%%%%%%%%%%%%%%%%
\subsection{The new variables.}

In a regular case, the coordinates in $TQ$ may be traded with those
in $T^*Q$ by means of the Legendre map, which is invertible. 
Instead, in a singular case (the gauge case), the Lagrangian definition
of the momenta, ${\hat p} := {\partial L}/{\partial {\dot q}}$ does not allow 
for the determination of $\dot q$ in terms of phase space coordinates. In 
fact, the definition of the momenta determines a certain number of independent
constraints 
$\phi_\mu(q,p)$, $\mu = 1,...,m$, in phase space, that is, functions such that 
their pullbacks $\phi_\mu(q,{\hat p})$ vanish identically. 
These functions locally define 
the primary constraint surface ${\cal M}_0 \in T^*Q$. Due to the presence of
these constraints, in order to trade the
coordinates in $TQ$ with those in this surface, we need to enlarge the phase 
space with $m$ new coordinates in $R^m$ (in fact, since all our results are 
local, all what we need is a open subset of $R^m$). For purposes that are made 
clear later on, we shall call $\lambda^\mu$ these new coordinates. 
The change of coordinates will be given by:
\bea
TQ & \longleftrightarrow & {\cal M}_0 \times R^m \\
q^i, \dot q^i & \longleftrightarrow & q^i,p_i,\lambda^\mu, \quad 
{\rm{} with} \quad \phi_\mu(q,p)=0,
\eea
with the following specific definition of one set of coordinates in terms 
of the other. For $ TQ  \longrightarrow {\cal M}_0 \times R^m $ we have
$$
p_i = {\hat p}_i(q, \dot q), 
\quad \lambda^\mu = v^\mu(q, \dot q ),
$$
with $v^\mu$ as defined in subsection 1.2; and for
$ TQ \longleftarrow  {\cal M}_0 \times R^m $, we have
\beq
\dot q^i = {\partial H_c\over\partial p_i} + 
\lambda^\mu {\partial \phi_\mu\over\partial p_i} 
= {\partial H_D\over\partial p_i}
\quad {\rm{} with} \quad \phi_\mu(q,p)=0.
\label{dotq}
\eeq

The complete change of variables includes higher order time derivatives 
corresponding to higher order tangent structures.
Thus, at the next level we have
\bea
\underline{\rm{old \, variables}} & \longleftrightarrow & 
\underline{\rm{new \, variables}} \\
q^i, \dot q^i, \ddot q^i \quad & \longleftrightarrow & q^i,p_i,\lambda^\mu,
\dot p_i, \dot \lambda^\mu,
\eea
with $\phi_\mu(q,p)=0 \ {\rm{} and} 
\  \dot \phi_\mu(q,p, \lambda, \dot p)=0$ for the new variables in the 
right hand side.
Observe that $\dot q$ does not appear in the right hand
 side because it is 
substituted by use of 
(\ref{dotq}). The new restrictions $ \dot \phi_\mu$ are nothing but the 
time derivatives of $\phi_\mu$, that can be written, 
using (\ref{dotq}), as 
$$
\dot  \phi_\mu(q,p, \lambda, \dot p) = \frac{\partial \phi_\mu}{\partial q}
{\partial H_D\over\partial p} +\dot p \frac{\partial \phi_\mu}{\partial p} .
$$

It is of great advantage to use, instead of $\dot p$, a new variable 
$l$ defined as
\beq
l := \dot p + {\partial H_D\over\partial q}.
\label{eles}
\eeq

When we undo the change of variables, we discover that $l = -[L] = 
-(\alpha - W \ddot q)$, that is, the Euler-Lagrange derivative of $L$. 
This means that setting $l=0$ will correspond to satisfying the Euler-Lagrange 
equations. Summing up, we have now the change of variables
\bea
\underline{\rm{old \, variables}} & \longleftrightarrow & 
\underline{\rm{new \, variables}} \\
q^i, \dot q^i, \ddot q^i & \longleftrightarrow & q^i,p_i,\lambda^\mu,
 l_i, \dot \lambda^\mu, \\ \quad 
{\rm{} with} \quad \phi^{(0)}_\mu(q,p):=\phi_\mu(q,p)=0 \ &{\rm{} and}& 
\  \phi^{(1)}_\mu(q,p, \lambda, l) :=\dot  \phi_\mu(q,p, \lambda, \dot p)
=0,
\eea
with the new relations
$$
\ddot q = \{ {\partial H_D\over\partial p}, H_D \} + 
 l {\partial H_D\over\partial p \,\partial p } + 
\dot \lambda {\partial H_D\over\partial p \,\partial \lambda } ,
$$
on one direction ($\leftarrow$), and 
$$l = -[L] = 
\dot {\hat p} - {\partial L \over\partial q}, \qquad \dot \lambda =\dot q {\partial v \over\partial q} 
+\ddot q {\partial v \over\partial \dot q}
$$
on the other($\rightarrow$). Restrictions $\phi^{(1)}_\mu$ can be written as 
\beq
\phi^{(1)}_\mu = \{\phi_\mu, \, H_D \} + l {\partial\phi_\mu\over\partial p},
\label{phi1}
\eeq
and they restrict the number of independent $l$ variables in the same way as 
$\phi^{(0)}_\mu :=\phi_\mu$ restricts the number of independent momenta. 

The next order is easily developed:
\bea
\underline{\rm{old \, variables}} & \longleftrightarrow & 
\underline{\rm{new \, variables}} \\
q^i, \dot q^i, \ddot q^i, {\dddot q}^i
& \longleftrightarrow & q^i,p_i,\lambda^\mu,
 l_i, \dot \lambda^\mu,  \dot l_i, \ddot \lambda^\mu,\\ \quad 
{\rm{} with} \quad \phi^{(0)}_\mu(q,p)=0,  
\quad & \phi^{(1)}_\mu(q,p, \lambda, l)=0,& \quad {\rm{} and} \quad 
 \phi^{(2)}_\mu(q,p, \lambda, l,\dot \lambda, \dot l )=0,
\eea
with  $\phi^{(2)}_\mu$ being the time derivative of $\phi^{(1)}_\mu$.
The change of coordinates at any order is defined along the same lines.
General formulas for $\phi^{(n)}$ will be given below.

The total time derivative operator, written as (\ref{timeder}) in 
terms of the 
old variables, takes with the new
variables the form:
\beq
\frac{d}{d\,t} = {\cal D} + l \frac{\partial}{\partial \, p} +  
 \dot l \frac{\partial}{\partial\,l} +  
\ddot l \frac{\partial}{\partial\, \dot l} + ... \ ,
\label{ntimeder}
\eeq
where ${\cal D}$ is defined as
\beq
{\cal D} := \frac{D \, }{D \, t} + \{ -, \, H_D \}
\label{cald}
\eeq
with ${D \, }/{D \, t}$ already defined in (\ref{Dt}).
But we must still bear in mind that the variables $p$, $l$, $\dot l$, etc., are
restricted by $\phi^{(0)}_\mu =0 ,\phi^{(1)}_\mu=0,
\phi^{(2)}_\mu=0,$ etc. We call these functions ``restrictions'', instead of
constraints because all them vanish identically when they are expressed in 
terms of the old variables. 

Notice the advantage of using the new variables in a non-regular 
theory: whereas with the old variables it is difficult to set the 
Euler-Lagrange equations to zero --i.e., to go on shell-- because these 
equations can not be written in normal form, 
it is trivial to do so when working with the 
new set variables --if one takes into account the restrictions appropriately.

%%%%%%%%%%%%%%%%%%%%%%%%%%%%%%%%%%%%%%%%%%%%%%%%%%%%%%%%%%%
%%%%%%%%%%%%%%%%%%%%%%%%%%%%%%%%%%%%%%%%%%%%%%%%%%%%%%%%%%%
%%%%%%%%%%%%%%%%%%%%%%%%%%%%%%%%%%%%%%%%%%%%%%%%%%%%%%%%%%%
\subsection{Restrictions for the new variables, general formulas.}

As it will be seen later, we do not need for our purposes the complete 
expressions for the restrictions
but only their expansion in terms the variables $l, \dot l, \ddot l, ... \ ,$ 
up to quadratic terms. The definition 
$$
\phi^{(n+1)}_\mu := \frac{d\,}{d\,t}(\phi^{(n)}),
$$
allows for a recursion formula that can be written as follows
\beq
\phi^{(n+1)}_\mu = {\cal D }^{n+1}\phi_\mu + \sum^n_{k=0} l^{(k)} \alpha(n,k)
\phi_\mu
+ ({\rm{} quadratic \ terms \ in}\ l, \dot l, \ddot l, ... \ ), 
\label{recurs}
\eeq
where ${\cal D }^{m}$ is the $m$ times composition of the operator 
(\ref{cald}), $\alpha(n,k)$ is a composition of operators
$$
\alpha(n,k) := \sum^{n-k}_{m=0} \left(\begin{array}{c} m+k \\ m 
\end{array} \right) {\cal D }^{m} \circ {\partial\over\partial p} \circ 
{\cal D }^{n-k-m},
$$
that satisfies, for $k>0$, the relation
$$
\alpha(n,k)+ \alpha(n,k-1)= \alpha(n+1,k),
$$
and  $l^{(k)}$ stands for the $k$-derivative of $l$, with $l^{(0)}=l$.

As we have said before, all these restrictions $\phi^{(n)}_\mu$ become 
identically zero when they are expressed in terms of the old variables.
However, if we set $l$, $\dot l$, $\ddot l$,..., to zero within 
$\phi^{(n)}_\mu$, we get constraints,
\beq
{\cal D }^{n}\phi_\mu \approx 0,
\label{constr}
\eeq 
that are not identically zero in
terms of the old variables ---except for $\phi_\mu$ itself. These constraints
will play a relevant role in our developments. 
We use Dirac's weak equality, 
$\approx$, for them because when viewed in terms of the
old variables $q^i, \dot q^i, \ddot q^i, {\dddot q}^i...$, they
are just combinations of the Euler-Lagrange equations and its derivatives. 
Thus, for $n=0$ we have $\phi_\mu(q, \hat p) =0$ identically; for $n=1$,
using (\ref {phi1}),
$$
({\cal D }^{1}\phi_\mu)(q, \hat p, v_\mu) = 
[L]_i{\partial\phi_\mu\over\partial p_i} = 
\alpha_i {\partial\phi_\mu\over\partial p_i},
$$
that are the primary Lagrangian constraints 
(Notice \cite{batlle86} that ${\partial\phi_\mu}/{\partial p_i}, 
\mu = 1 \cdots m $ form a basis for the
 null vectors of the Hessian matrix
(\ref{hess})), and so on.
%%%%%%%%%%%%%%%%%%%%%%%%%%%%%%%%%%%%%%%%%%%%%%%%%%%%%%
%%%%%%%%%%%%%%%%%%%%%%%%%%%%%%%%%%%%%%%%%%%%%%%%%%%%%%
\subsection{Relation with Dirac's constraints.}

Constraints (\ref{constr}) are not in the form of Dirac constraints, as
obtained in Dirac's stabilization algorithm for constrained system 
\cite{dirac},  \cite{batlle86}, \cite{sudh-muk},  but they have the 
same content. Dirac's
algorithm is more refined than what we need here, and is able to 
reformulate the whole set of constraints (\ref{constr}) as {\it a)} some 
standard Dirac
constraints of the type $\psi(q,p)$, and {\it b)} the determination of some of 
the Lagrange multipliers $\lambda$ as functions in phase space. 
Dirac's clever trick relies
in the classification of constraints as first and second class. Let
us see how it works for our ``secondary'' constraints ${\cal D }\phi_\mu$.

If we distinguish, among the primary constraints $\phi_\mu$, those
that are first class, $\phi_{\mu_0}$, and those second class,  $\phi_{\mu_0'}$,
then
$$
{\cal D }\phi_{\mu_0} = \{\phi_{\mu_0}, \, H_c\} + 
\lambda^\nu \{\phi_{\mu_0},\, \phi_\nu \},
$$
and since the first class condition makes the second piece, 
$\{\phi_{\mu_0},\, \phi_\nu \}$, to vanish in the
surface of the primary constraints, we are left with the secondary Dirac 
constraints 
$$
\phi^1_{\mu_0}:=\{\phi_{\mu_0}, \, H_c\}.
$$

On the other hand, the requirement
$$
{\cal D }\phi_{\mu_0'} = \{\phi_{\mu_0'}, \, H_c\} + 
\lambda^\nu \{\phi_{\mu_0'},\, \phi_\nu \} =0,
$$
allows for the canonical determination of $\lambda^{\nu_0'}$ as a function
in phase space, 
$\lambda^{\nu_0'} =\lambda_c^{\nu_0'}(q,p)$, because
the matrix $\{\phi_{\mu_0'},\, \phi_{\nu_0'}\}$ is regular as implied by the 
second class condition. In the standard Dirac's method, this determination
of some of the variables $\lambda$ is then introduced into the dynamics and
the operator ${\cal D }$ is modified to
$$
{\cal D' }  := \frac{D' \, }{D \, t} + \{ -, \, H'_D \}
$$
where
$H'_D = H'_c + \lambda^{\nu_0}\phi_{\mu_0}$ and
$H'_c = H_c + \lambda_c^{\nu_0'}\phi_{\mu_0'}$. Now $\frac{D' \, }{D \, t}$ 
is the adaptation of equation (\ref{Dt}) to the Lagrange multipliers that are 
left undetermined, that is, $\lambda^{\nu_0}$.

Once the dynamics has been adapted to the partial knowledge of the
Lagrange multipliers, we only need to care about the new constraints
$\phi^1_{\mu_0}$ and require ${\cal D'}\phi^1_{\mu_0} =0$ as a new
set of constraints. Then the whole mechanism starts again. This is Dirac 
method. For what the constraints (in the enlarged formalism) of the form 
$\lambda^{\nu_0'} - \lambda_c^{\nu_0'}(q,p)=0$ are concerned, the
application of the new evolution operator ${\cal D'}$
leads to new constraints involving the time derivative of $\lambda^{\nu_0'}$,
that is,
$$
\dot \lambda^{\nu_0'} - \{\lambda_c^{\nu_0'}(q,p) , \, H'_D \} = 0,
$$
and so on. This information is irrelevant from the point of view of the
Dirac method because the variables $\lambda^{\nu_0'}$ have been 
substituted by their canonical determinations $\lambda_c^{\nu_0'}$ and thus 
have disappeared from the formalism.

%%%%%%%%%%%%%%%%%%%%%%%%%%%%%%%%%%%%%%%%%%%%%%%%%%%%%%%

\section{Reformulation of the Noether condition with the new variables.}

Consider a theory defined by a Lagrangian $L$ and with a Noether transformation
$\delta^L q$ of the type (\ref{gen-delq}) for which there exists a function 
$G^L(q,\, \dot q, \ddot q, ... ;\,t)$ satisfying (\ref{noet}). 
Now we express $\delta^L q$ and $G^L$ in terms of 
the new variables. We end up
with (for definiteness we fix the highest order arguments, the superscript $E$
stands for reference to the enlarged formalism)
\beq
\delta^E q(q,p,\lambda,
 l, \dot \lambda,  \dot l, \ddot \lambda, ...,
l^{(f+1)}, \lambda^{(f+2)}),   \quad 
  G^E(q,p,\lambda,
 l, \dot \lambda,  \dot l, \ddot \lambda, ...,
l^{(f)}, \lambda^{(f+1)}),
\label{objects}
\eeq
such that 
\beq
\delta^L q(q,\, \dot q, \ddot q, ... ;\,t) =
\delta^E q(q,{\hat p}, v^\mu, -[L], \frac{d^L\,v^\mu}{d\,t}, ...;t), 
\label{delq-le}
\eeq
\beq
G^L(q,\, \dot q, \ddot q, ... ;\,t) =
G^E(q,{\hat p},v^\mu, -[L], \frac{d^L\,v^\mu}{d\,t}, ...;t).
\label{g-le}
\eeq

Notice that $\delta^E q$ and $G^E$ are not uniquely defined, for we can 
add to them any linear combination of the restrictions $\phi^{(n)}_\mu$ 
with no effect in $\delta^L q$ and $G^L$. These 
arbitrariness will be fixed below at their
 due moment. Expression 
(\ref{noet}) now becomes, in terms of the new variables,
$$
-l \delta^E q + \frac{d\,G^E}{d\,t} \sota{restr} 0
$$ 
where $\sota{restr}$ stands for an equality under the restrictions 
$\phi^{(n)}_\mu = 0, \, n = 0, 1, ...f+2$.  
According to the theory of Lagrange multipliers, this means that 
there exist functions $a_n^\mu$ such that 
\beq
-l \delta^E q + \frac{d\,G^E}{d\,t} = \sum^{f+2}_{n=0} a_n^\mu \phi^{(n)}_\mu,
\label{newnoet}
\eeq 
with all the variables now taken as independent. 

Notice that, for a given $n > 0$, 
\beq
a^\mu_n \phi_\mu^{(n)} = a^\mu_n \frac{d\,}{d\,t} \phi_\mu^{(n-1)} = 
\frac{d\,}{d\,t}
\left(a^\mu_n \phi_\mu^{(n-1)} \right) - \left(\frac{d\,}{d\,t}a^\mu_n\right) 
\phi_\mu^{(n-1)}.
\label{recur}
\eeq
Therefore, if we define a new set of functions, 
$$
b_m^\mu = - \sum_{n=m+1}^{f+2}(-1)^{n-m-1}\ a_n^{\mu  (n-m-1)},
$$ 
for $m= 0, \cdots, f+1$ and with $a_n^{\mu (k)}$ being the $k$-th time 
derivative of $a_n^{\mu}$, then (\ref{newnoet}) can be rewritten as
$$
-l \delta^E q + \frac{d\,}{d\,t} \left(G^E+\sum^{f+1}_{m=0} 
b^\mu_m \phi_\mu^{(m)}\right) = pc.
$$
Here $pc$ stands, as usual, for a linear combination
of the primary constraints $\phi_\mu$ that define the surface 
${\cal M}_0 \in T^*Q$.
 Now we take advantage of the arbitrariness of the choice
of $G^E$ under the addition of pieces linear in the restrictions 
$\phi_\mu^{(n)}$ to absorb the piece 
$\sum^{f+1}_{m=0}b^\mu_m \phi_\mu^{(m)}$ into $G^E$.
We can conclude that a function $G^E$ exists, satisfying (\ref{g-le}), such 
that, together with a given function $\delta^Eq$ satisfying (\ref{delq-le}), 
the following relation holds: 
\beq
-l \delta^E q + \frac{d\,}{d\,t} G^E = pc.
\label{newnoet2}
\eeq
Notice that (\ref{newnoet2}) reduces to (\ref{enl-cond}) 
when we enforce $l=0$ into it.
Let us expand (\ref{newnoet2}) up to quadratic terms in $l$ and
its time derivatives. Defining the expansions for 
$\delta^E q \ {\rm{} and}\ G^E$ as
\bea
\label{expq}
\delta^E q &=& \delta_0 q + 
({\rm{}linear \, terms \, in \,} l, \dot l, ...), \\
G^E &=& G^c + \sum^{f}_{k=0} l^{(k)} G_k +  ({\rm{} quadratic \, terms\, 
in \,} l, \dot l, ...), 
\label{expg}
\eea
(now $\delta_0 q, G^c, {\rm{} and} \, G_k$ only depend on 
$(q, p, \lambda, \dot \lambda, \ddot \lambda...)$, that is, they do not 
depend on the variables $l^{(k)}$) we get, for (\ref{newnoet2}),
\beq
-l \delta_0 q + \frac{d\,}{d\,t} (G^c 
+\sum^{f}_{k=0}l^{(k)} G_k) + 
({\rm{}quadratic \, terms \, in \,} l, \dot l, ...)= pc. 
\label{newnoet3}
\eeq
Using (\ref{ntimeder}), (\ref{newnoet3}) gives
\beq
-l \delta_0 q + {\cal D}G^c + l \frac{\partial\,G^c}{\partial\,p} 
+\sum^{f+1}_{k=0}l^{(k)} ({\cal D}G_k +G_{k-1}) + 
({\rm{}quadratic \, terms \, in \,} l, \dot l, ...)= pc, 
\label{newnoet4}
\eeq
with the new definitions $G_{f+1} := 0$ and $G_{-1} := 0$. 
Now, isolating in (\ref{newnoet4}) the pieces with no dependence on $l$, we get
$$
{\cal D}G^c = pc,
$$
which is nothing but the condition (\ref{enl-cond}) for a function 
$G^c(q, p, \lambda, \dot \lambda, \ddot \lambda...)$ to
be a canonical generator of a Noether transformation within the enlarged 
formalism. The expansion for $l^{(k)}$ in (\ref{newnoet4}) gives, for $k>0$,
$$
{\cal D}G_k +G_{k-1} = c^\mu_k \phi_\mu,
$$
for some functions $c^\mu_k$. Notice then that this result, together with 
the identity 
$$
G_0 = \sum^{f}_{k=0}(-1)^{(k)}{\cal D}^k({\cal D}G_{k+1} +G_k),
$$
allows to write $G_0$ as 
$$
G_0 =\sum^{f}_{k=0}(-1)^{(k)}{\cal D}^{k}(c^\mu_{k+1} \phi_\mu), 
$$
that is, $G_0$ is a combination of the constraints ${\cal D}^{k} \phi_\mu$ 
introduced in (\ref{constr}). Finally, the expansion for $l$ in 
(\ref{newnoet4}) gives
$$
-\delta_0 q + \frac{\partial\,G^c}{\partial\,p}+{\cal D}G_0 = c^\mu_0 \phi_\mu,
$$
for some functions $c^\mu_0$. But since $G_0$ is a combination of 
constraints, so it is ${\cal D}G_0$, and therefore we arrive at 
\beq
\delta_0 q = \frac{\partial\,G^c}{\partial\,p} - 
\sum_{n=0}^{f+1} d^\mu_n{\cal D}^n \phi_\mu,\, 
\label{dq0}
\eeq
for some functions $d^\mu_n$. Now, as we did before with $G^E$, we can 
use the arbitrariness that equation (\ref{delq-le}) allows for the 
choice of $\delta^Eq$ and add to it a linear combination of
the restrictions, namely $d^\mu_n\phi^{(n)}_\mu$. 
The new $\delta_0 q$ deduced from the expansion (\ref{expq}) for the new
$\delta^Eq$
will absorb the piece $d^\mu_n\phi^{(n)}_\mu|_{l^{(k)}=0} = 
d^\mu_n{\cal D}^n \phi_\mu$ in (\ref{dq0}) and therefore it will satisfy,  
\beq
\delta_0 q = \frac{\partial\,G^c}{\partial\,p} = \{q, \, G^c \}.
\label{thersult}
\eeq

Equation (\ref{thersult}) is the main result of our paper. 
To summarize, we have considered, as a starting point, a Noether 
transformation $\delta^Lq$ 
--satisfying (\ref{dell})-- of the most general type (\ref{gen-delq}), 
and we have obtained the objects $\delta_0 q$ and
$G^c$ in the enlarged formalism such that they satisfy the Noether
condition (\ref{enl-cond}) and such that $\delta_0 q$ is canonically generated
by $G^c$, according to (\ref{thersult}). The results of \cite{noeth98}
guarantee that $\delta_0 q|_{\lambda^\mu =v^\mu(q, \dot q)}$ 
\footnote{Substitution of $\dot\lambda, \ddot \lambda,...$ is also 
understood in this kind of expressions.} 
is a Noether transformation for the Lagrangian $L$.

Taking into account equations (\ref{delq-le}) and (\ref{g-le}), and the
expansions (\ref{expq}) and (\ref{expg}), we observe that this new
Noether transformation 
$\delta_0 q|_{p=\hat p,\,\lambda^\mu =v^\mu(q, \dot q)}$ 
differs from the original one, $\delta^Lq$, at most by a
combination of the equations of motion, $[L]$, and its time derivatives. 
They coincide {\it on-shell}. The same is true for the relation between 
$G^L$ and $G^c|_{p=\hat p,\,\lambda^\mu =v^\mu(q, \dot q)}$.

\vspace{4mm}

Our conclusion can be stated as follows:
\begin{teor}  
Given a first order Lagrangian $L(q,\dot q)$, 
and a generalized Noether transformation for it,
$$ 
\delta^L q (q,\, \dot q, \ddot q, ... ;\,t),
$$
with its associated conserved quantity 
$$
G^L(q,\, \dot q, \ddot q, ... ;\,t),
$$
there always exists a Noether transformation, 
$$
\delta_0 q (q,p,\lambda, \dot \lambda, \ddot \lambda... ;\,t),
$$
and a conserved quantity, 
$$
G^c(q,p,\lambda, \dot \lambda, \ddot \lambda... ;\,t),
$$
in the enlarged formalism (with Lagrangian $L_c$) such that 
$$
\delta_0 q = \frac{\partial\,G^c}{\partial\,p},
$$
and that 
$$
\delta_0^L q 
:= \delta_0 q|_{[p=\hat p,\,\lambda^\mu =v^\mu(q, \dot q)]},
$$
and 
$$
G_0^L := G^c|_{[p=\hat p,\,\lambda^\mu =v^\mu(q, \dot q)]},
$$
are a Noether transformation and its associated conserved quantity for the 
Lagrangian $L$, and that they differ from $\delta^L q$ and $G^L$, 
respectively, by terms that are,
at most, a combination of the equations of motion and its time derivatives 
--that is, they coincide on shell.
\end{teor}

This theorem can be summarized as follows:

{\sl Any Noether symmetry $\delta^L q (q,\, \dot q, \ddot q, ... ;\,t)$ 
can be expressed, up to terms that vanish on-shell, as a canonical 
transformation in a phase space enlarged with the variables 
$\lambda, \dot \lambda, \ddot \lambda , \dots $, where $\lambda$ are 
the Lagrange 
multipliers associated with the primary constraints.}

%%%%%%%%%%%%%%%%%%%%%%%%%%%%%%%%%%%%%%%%%%%%%%%%%%%%%%%%%
%%%%%%%%%%%%%%%%%%%%%%%%%%%%%%%%%%%%%%%%%%%%%%%%%%%%%%%%%
\section{Application to a regular theory.}

Now we can give an answer to the question raised at the beginning
of section 2. 
In the case of a regular theory, that is, when the Hessian (\ref{hess})
is regular, there are no constraints and hence the variables $\lambda$
in the enlarged formalism do not appear. This means that $\delta_0q$ will
be only $\delta_0q(q,p;t)$ and $G^c$ will be $G^c(q,p;t)$. Since the
Euler-Lagrange equations $[L]=0$ allow for isolating
\beq
\ddot q^i = (W^{(-1)})^{ij}\alpha_j,
\label{aill}
\eeq
what we find is that 
$$
\delta^Lq|_{\ddot q = (W^{(-1)})\alpha} = \delta_0q|_{p=\hat p},
$$
where it is understood that the higher time derivatives in $\delta^Lq$ are
also substituted by lower ones by the repeated use of (\ref{aill}) and its
time derivatives.

It is also true that, in this case, 
$$
G^L|_{\ddot q = (W^{(-1)})\alpha} = G^c|_{p=\hat p}.
$$

\vspace{4mm}

We can conclude that {\it in a regular theory defined by a first order 
Lagrangian, the most general
Noether transformation coincides {\it on-shell} with a Noether
transformation that is canonically generated in phase space.}
%%%%%%%%%%%%%%%%%%%%%%%%%%%%%%%%%%%%%%%%%%%%%%%%%%%%%%%%%%%%%%%%%%%
%%%%%%%%%%%%%%%%%%%%%%%%%%%%%%%%%%%%%%%%%%%%%%%%%%%%%%%%%%%%%%%%%%%
\section{Revisiting the Noether identities.}

According to the results of the previous sections, the most general continuous
Noether transformations for a given Lagrangian is associated, up to 
constraints of the formalism, to a generator  
$G(q,p,\lambda, \dot \lambda, \ddot \lambda... ;\,t)$ ($G^c$ in section 3)
that satisfies (\ref{enl-cond}),
$$
{\cal D}G = pc.
$$
Consider now the particular case of a gauge transformation, this means that
$G$ can be expanded as
$$
G=\sum^M_{i=0} \epsilon^{(i)}G_i,
$$
with $\epsilon^{(0)}$ an arbitrary function and 
$$
\epsilon^{(i+1)} = \frac{d\,}{d\,t} (\epsilon^{(i)}).
$$
Plugging this expansion into (\ref{enl-cond}) we get
\bea
G_M &=& pc, \\
G_{i-1} + {\cal D}G_i &=& pc , \ \  i = 1, \cdots , M  \\ 
{\cal D}G_0 &=& pc.
\eea
If we define $G_{M+1} = G_{-1} = 0$, this expression 
can be summarized as $G_{i-1} + {\cal D}G_i = pc$  for  
$i = 0, \cdots , M+1$. Then, taking into account (\ref{ntimeder}), we can
define the quantities $K_i$ as
\beq
K_i :=  G_{i-1} + \frac{d\,}{d\,t} (G_i) - 
l \frac{\partial G_i}{\partial \, p} = pc = e^\mu \phi_\mu,  
\label{thek}
\eeq
for some coefficients $e^\mu$.
The quantities $K_i$ are therefore primary constraints. 
Using (\ref{thek}), we can construct the following relation
\bea
\nonumber \sum^{M+1}_{i=0} (-1)^i 
\left(\frac{d\,}{d\,t}\right)^i (K_i) &=&
\sum^{M+1}_{i=0} (-1)^i \left(\frac{d\,}{d\,t}\right)^i(G_{i-1} 
+ \frac{d\,}{d\,t} (G_i)) 
 \\
&-& \sum^{M}_{i=0} (-1)^i \left(\frac{d\,}{d\,t}\right)^i 
(l \frac{\partial G_i}{\partial \, p}).
\label{thedk}
\eea
Since the first piece in the right hand side of (\ref{thedk}),
vanishes identically, use of (\ref{thek}) yields
$$
\sum^{M}_{i=0} (-1)^i \left(\frac{d\,}{d\,t}\right)^i 
(l \frac{\partial G_i}{\partial \, p}) = e^\mu_n \phi_\mu^{(n)},
$$
for some coefficients $e^\mu_n$.
Now we can take advantage of the fact that in terms of the old variables,
the restrictions 
$\phi_\mu^{(n)}$ vanish identically and the variables  $l$ become the 
Euler-Lagrangian equations $-[L]$, to get
\beq
\sum^{M}_{i=0} (-1)^i \left(\frac{d^L\,}{d\,t}\right)^i ([L] f_i) = 0,
\label{neotherident}
\eeq 
identically, where
$$
f_i := \frac{\partial G_i}{\partial \, p}
|_{p=\hat p,\,\lambda^\mu =v^\mu(q, \dot q)}.
$$

Equation (\ref{neotherident})
is the Noether identity corresponding to the gauge transformation 
$$
\delta q =
  \frac{\partial G}{\partial \, p}=
\sum^M_{i=0} \epsilon^{(i)}f_i.
$$

%%%%%%%%%%%%%%%%%%%%%%%%%%%%%%%%%%%%%%%%%%%%%%%%%%%%%%%%%%%%%%%
%%%%%%%%%%%%%%%%%%%%%%%%%%%%%%%%%%%%%%%%%%%%%%%%%%%%%%%%%%%%%%%
 
\section{Example}
Here we will consider as an example an equivalent formulation of 
the conformal particle, first 
introduced in \cite{marnelius} and developed in \cite{siegel}. This system 
has attracted 
attention recently \cite{bars}  by gauging the $Sp$(2,R) invariance
of a zero Hamiltonian system whose  canonical Lagrangian  gives an 
equivalent way to define the conformal particle. It is then possible to write 
down
an action that have interesting properties like rigid $SO$(2,$D$) symmetry
and a rich structure in the reduced phase space. This properties has
been used recently in connection with string theory \cite{bars1}, $AdS$ space 
time and in connection to the so called M-theory \cite{bars2}. As another 
application the system can
be used for isometric embedding of BPS branes in flat spaces with two 
times \cite{proeyen}.
Here we will consider the equivalent system defined by the Lagrangian
\beq
L=\frac{\dot X^2}{2e}-\frac12 X^2,
\label{lag-ex}
\eeq
in the configuration space defined by the variables $(e,X^M)$, $M=0,1,2,...D$
in a space-time with ``two times'' with metric 
$\eta=\mbox{diag}(-1,-1,1,...1)$.
These ``two times'' are necessary to solve the constraints without ghosts.
A generalized Lagrangian symmetry for this system is 
\beq
\delta^L X=-2\Bigg(\frac{\dot X}{e}\Bigg)^{\cdot\cdot}\epsilon
-\Bigg(\frac{\dot X}{e}\Bigg)^{\cdot}\dot\epsilon+
\frac{\dot X}{e}\ddot\epsilon,
\label{sym-ex}
\eeq
and 
$$
\delta^L e=\dddot \epsilon + 4e\dot\epsilon+2\dot e \epsilon,
$$ 
whose associated Lagrangian generator is
\beq
G^L=-\epsilon
(\frac1e \ddot X
+X-\frac{1}{e^2}\dot e \dot X)^2
 +(\ddot\epsilon+2\epsilon e) \frac{\dot X^2}{2e^2}+
\frac{\dot\epsilon}{e}X\cdot\dot X+ \epsilon X^2,
\label{gen-ex}
\eeq
where $\epsilon$ is an arbitrary parameter and the dot is a short notation for
$\frac{d^L}{dt}$ (see eq. (\ref{timeder})). This generator, $G^L$, and
 its associated symmetries $\delta^L e$, $\delta^L X$ satisfy (\ref{noet}).
 Notice that (\ref{sym-ex}) and (\ref{gen-ex}) depends on higher order
derivatives of the variables $X^M$ with respect to time. 
Nevertheless, we will show that it is possible to associate this 
generalized Lagrangian symmetry with a canonical symmetry 
in the enlarged phase space (containing the 
Lagrange multipliers as variables). In turn, this 
canonical symmetry 
generates the correct $Sp$(2,R) symmetry in the sector $(X^M,P^M)$ of the 
initial conditions in the enlarged phase space. 

To construct the  associated canonical formalism for the Lagrangian 
(\ref{lag-ex}) we start from the definition of the canonical momenta
$$P_e=0,\quad P=\frac{\dot X}{e},$$
so, $P_e=0$ is a primary constraint. The Dirac Hamiltonian is
$$H_D=\frac12 e P^2+\frac12 X^2+\lambda P_e,$$ 
where $\lambda$ is the Lagrange
multiplier associated with the primary constraint $P_e=0$ whose 
successive stabilization produces the Dirac first class constraints
$P^2=0$, $X\cdot P=0$ and $X^2=0$ that close under $Sp$(2,R) Lie algebra. 
To associate with the given Lagrangian symmetry a symmetry 
in the enlarged space we will use the change of variables introduced 
in section 2. 
The dictionary is given by 
$$ \dot e= \lambda, \quad \dot X=eP,$$
$$\ddot e=\dot\lambda, \quad \ddot X=\lambda P+e(-X+l_X),$$
$$\dddot e=\ddot\lambda, \quad \dddot X=\dot\lambda P+2\lambda(-X+l_X)
 +e(-eP+\dot l_X),$$
where 
$$l_X=\dot P+\frac{\partial H_D}{\partial X},$$
(see eq. (\ref{eles})). Using this change of variables we construct 
the objects (\ref{objects}), that satisfy    
(\ref{delq-le}) and (\ref{g-le}) and, consequently, (\ref{newnoet}). With 
the appropriate addition of restrictions we can prepare $\delta^E q$ and 
$G^E$ to satisfy (\ref{newnoet2}). The result is 
\beq
G_E=G^c-\epsilon l_X^2,
\label{enl-gen-ex}
\eeq
with
$$
G^c=(\dddot\epsilon+4\dot\epsilon e -2\epsilon\lambda)P_e+\frac12(\ddot\epsilon
+2\epsilon e)P^2+\dot\epsilon X\cdot P+\epsilon X^2,
$$
and
$$
\delta^E X=(\ddot \epsilon +2\epsilon e) P+\dot\epsilon X-
2\epsilon\dot l_X-\dot\epsilon l_X,
$$
$$
\delta^E e=\dddot\epsilon+4e\dot\epsilon+2\lambda\epsilon.
$$

Now we can apply our theorem as stated in section 3 by noticing
that the sector that does not depends on  $l_X$ in (\ref{enl-gen-ex}) 
generates a canonical symmetry that will reduce to the original 
symmetry in the $n$-th tangent space up to constraints. Indeed,
$$
\delta_{0} X=\{X,G^c\}=(\ddot \epsilon +2\epsilon e) P+\dot\epsilon X,
$$
$$
\delta_{0} e=\{e,G^c\} =\dddot\epsilon+4\dot\epsilon e 
+ 2\epsilon\lambda,$$
is a canonical symmetry for the system under consideration. 
This can be checked explicitly by noticing that $G^c$ solves 
(\ref{enl-cond}) in the phase space.
When this symmetry is displayed in the initial conditions surface, the 
arbitrary function $\epsilon$ and its derivatives $\dot\epsilon$, 
$\ddot\epsilon$, $\dddot\epsilon$ become all of them arbitrary parameters. 
The gauge symmetry contains then three independent generators that 
form an $Sp$(2,R) algebra in the $(X,P)$ sector of 
phase space that corresponds to the algebra of the first class constraints 
$P^2, P\cdot X$ and $X^2$, as expected.

For completeness we  list the  restrictions associated with this 
problem and the Noether identity that follows from them. The first four
restrictions are
\bea
\phi^{(0)} &=& P_e, \\
\phi^{(1)} &=& -\frac12 P^2 +l_e=0,\\
\phi^{(2)} &=& X\cdot P-P\cdot l_X+\dot l_e,\\
\phi^{(3)} &=& e P^2-X^2+2X\cdot l_X-l^2_X-P\cdot\dot l_X+\ddot l_e.
\eea

Notice that the Dirac constraints can be obtained from these restrictions
by enforcing $l_e$ and $l_X$ to zero. The Noether identity is 
$$
2 \dot e [L]_e+4e\dot {[L]}_e+3X\cdot\dot {[L]}_X-3\dot {[L]}_X\cdot {[L]}_X
-{ \dot X \over e} \cdot\ddot {[L]}_X +\dddot {[L]}_e=0.
$$
This identity can be obtained as follows. First we get the 
stabilization of the last restriction under the evolution 
operator (\ref{ntimeder}), and then we substitute the terms that contain 
Dirac constraints by use of the previous restrictions and taking into 
account that $l_X=-[L]_X$, $l_e=-[L]_e$ and $\lambda=\dot e$.
Another way to do the same is by noticing that the
combination of restrictions given by
$\phi^{(4)} + 2 \lambda \phi^{(1)} + 4 e \phi^{(2)}$, is precisely the
Noether identity just mentioned, after performing the substitution 
$l_X=-[L]_X$, $l_e=-[L]_e$ and $\lambda=\dot e$.  
Notice that the Lagrangian generalized symmetry
that we started from can be constructed by using this Noether identity.

\section{Appendix}

Let us establish the 
conditions for a function $G^c(q,p,\lambda, \dot \lambda, \ddot 
\lambda,...;t)$ to be a Noether generator for the Lagrangian $L_c$, 
under the definitions
\begin{equation}
\delta_c q ^{i} = \{q^{i}, \, G^c \}, \quad \delta_c p_{i} = \{p_{i}, 
\, G^c \},
\end{equation}
and with $\delta_c \lambda^\mu$ to be determined below. 

Compute $\delta_c L_c$,
\begin{eqnarray}
\nonumber
\delta_c L_c & = & \delta_c p_i \,{\dot q^i} + {d\over dt}{( p_i 
\,\delta_c q^i)} -
{\dot p_i} \delta_c q^i - \delta_c H_c
- \lambda^\mu \delta_c \phi_\mu - (\delta_c \lambda^\mu) \phi_\mu \\ 
\nonumber
& = & {d\over dt}{( p_i \,\delta_c q^i)} + \{p_i,\, G^c \} {\dot q^i} 
- \{q^i,\, G^c
\} {\dot p_i}\\ \nonumber 
&-& \{ H_c,\, G^c \}- \lambda^\mu \{ \phi_\mu,\, G^c \} 
- (\delta_c \lambda^\mu) 
\phi_\mu\\ \nonumber
& = & {d\over dt}{( p_i \,\delta_c q^i)} - {\dot q^i} \frac{\partial 
G^c} {\partial q^i} - {\dot p_i} \frac{\partial G^c}{\partial p_i} - 
\{ H_c,\, G^c \}
- \lambda^\mu \{ \phi_\mu,\, G^c \} - (\delta_c \lambda^\mu) 
\phi_\mu\\ \nonumber
& = & {d\over dt}{( p_i \,\delta_c q^i)} - \frac{d G^c}{d \, t} + 
\frac{\partial G^c}{\partial t}
+ {\dot \lambda^\mu} \frac{\partial G^c}{\partial \lambda^\mu} + 
{\ddot \lambda^\mu} \frac{\partial G^c}{\partial \dot {\lambda^\mu}} 
+ ...\\
\nonumber
&& \, + \, \{ G^c,\, H_c \}
- \lambda^\mu \{ \phi_\mu,\, G^c \} - (\delta_c \lambda^\mu) \phi_\mu 
\\ \nonumber
& = & {d\over dt}{( p_i \,\delta_c q^i - G^c) } + \frac{D G^c}{D t}
+ \{ G^c,\, H_D \}
- (\delta_c \lambda^\mu) \phi_\mu.
\end{eqnarray}

If we require
\begin{equation}
 {\cal D}G^c := \frac{D G^c}{D t} + \{ G^c,\, H_D \} = pc, 
\label{enl-cond2}
\end{equation}
and if we represent this combination $pc$ of primary constraints as 
$pc = C^\mu \phi_\mu$, then the definition \begin{equation}
\delta_c \lambda^\mu = C^\mu,
\label{dellamb}
\end{equation}
makes
$\delta_c L_c = {d\over dt}{( p \,\delta_c q - G^c) }$, that is, a 
Noether transformation for the enlarged formalism. (\ref{enl-cond2}) is the 
result (\ref{enl-cond}) we were looking for.
%%%%%%%%%%%%%%%%%%%%%%%%%%%%%%%%%%%%%%%%%%%%%%%%%%%%%%%%%
%%%%%%%%%%%%%%%%%%%%%%%%%%%%%%%%%%%%%%%%%%%%%%%%%%%%%%%%%

\end{document}